\documentclass{article}

\usepackage{amssymb}

\title{Sequent Calculus for Intuitionistic Epistemic Logic}
\author{Vladimir N.~Krupski \ and Alexey Yatmanov \\[6pt]
{\small Faculty of Mechanics and Mathematics,} \\
{\small Lomonosov Moscow State University, Moscow
119992, Russia.} \\
{\small \tt krupski@lpcs.math.msu.su} }

\date{}

\usepackage{prooftree}

\def\ruleone#1#2{\prooftree #1 \justifies #2 \endprooftree}
\def\ruletwo#1#2#3{\ruleone{#1\quad #2}{#3}}

\def\To{\Rightarrow}

\newtheorem{theo}{Theorem}[section]
\newtheorem{lem}[theo]{Lemma}
\newtheorem{cor}[theo]{Colorrary}
\newtheorem{defin}[theo]{Definition}

\def\proof{\par\smallskip\noindent {\bf Proof. }}
\def\com{\par\medskip\noindent{\bf Comment. }}
\def\fin{\hfill\rule{1ex}{1ex}\\[2pt]}

\def\IEL{{\sf IEL}}
\def\IELG{{\sf IEL_{G}^0}}
\def\IELGm{{\sf IEL_{G}^-}}
\def\IELGp{{\sf IEL_{G}}}
\def\IELGpp{{\sf IEL_{G}'}}

\begin{document}
\maketitle

\begin{abstract} The formal system of intuitionistic epistemic logic \IEL \ was
proposed by S.~Artemov and T.~Protopopescu. It provides the formal foundation for the
study of knowledge from an intuitionistic point of view based on
Brouwer-Hayting-Kolmogorov semantics of intuitionism. We construct a cut-free sequent
calculus for \IEL \ and establish that polynomial space is sufficient for the proof
search in it. So, we prove that \IEL \ is {\sf PSPACE}-complete.
\end{abstract}

\section{Introduction}
Modal logic \IEL, the basic Intuitionistic Epistemic Logic, was proposed by S.~Artemov
and T.~Protopopescu in \cite{ArtProt}. It was defined  by the following calculus.

Axioms:
\begin{itemize}
\item Axioms of propositional intuitionistic logic

\item $K(F\to G)\to (KF\to KG)$ \quad (distribution)

\item $F\to KF$ \quad(co-reflection)

\item $\neg K\bot$ \quad(consistency)
\end{itemize}

Rule: $F,\,F\to G\vdash G$ (Modus Ponens)

\medskip
Here knowledge modality $K$ means verified truth, as suggested by  T.~William\-son in
\cite{Will}. According to the Brouwer-Heyting-Kolmogorov semantics of intuitionistic
logic, a proposition is true iff it is proved. The co-reflection principle states that
any such proof can be verified.

The intuitionistic meaning of implication provides an effective proof checking procedure
that produces a proof of $KF$ given a proof of $F$. But the assumption that its output
always contains a proof of $F$ is too restrictive. The procedure may involve some
trusted sources which do not necessarily produce explicit proofs of what they
verify\footnote{ For example, it can be some trusted database that stores true facts
without proofs or some zero-knowledge proof.}. So the backward implication which is the
reflection principle $KF\to F$ used in the classical epistemic logic (see
\cite{FaHaMoVa}) is wrong in the intuitionistic setting. In general, a proof of $KF$ is
less informative than a proof of $F$.

At the same time some instances of the reflection principle are true in \IEL. In
particular, it is the consistency principle which is equivalent to $K\bot\to\bot$. The
proof of $K\bot$ contains the same information as the proof of $\bot$ because there is
no such proof at all. The more general example is the reflection principle for negative
formulas: $K\neg F\to\neg F$. It is provable in \IEL \ (see \cite{ArtProt}).

In this paper we develop the proof theory for \IEL. Our main contributions are the
cut-free sequent formulation and the complexity bound for this logic. It is established
that polynomial space is sufficient for the proof search, so \IEL \ is PSPACE-complete.

Our cut-elimination technique is syntactic (see \cite{TrSh}). We formulate a special
cut-free sequent calculus $\IELGm$ without structural rules (see Section \ref{SEC3})
that is correct with respect to the natural translation into \IEL. It has a specific
$K$-introduction rule $(KI_1)$ that also allows to contract a formula $F$ in the
presence of $KF$ in antecedents. This choice makes it possible to prove the
admissibility of the standard contraction rule as well as the admissibility of all
natural \IEL-correct modal rules (Sections \ref{SEC4}, \ref{SEC5}). The admissibility of
the cut-rule is proved by the usual induction on the cutrank (Section \ref{SEC6}). As
the result we obtain the equivalence between $\IELGm$ and $\IELG$. (The latter is the
straightforwardly formulated sequent counterpart for \IEL \ with the cut-rule.) Finally
we formulate a light cut-free variant of $\IELGm$ with the contraction rule and with
modal rules
$$
\ruleone{\Gamma_1,\Gamma_2\To F} {\Gamma_1,K(\Gamma_2)\To KF\using{(KI)}}\,,
\qquad\qquad \ruleone{\Gamma\To K\bot}{\Gamma\To F\using{(U)}}\,.
$$
It is equivalent to $\IELGm$.

The proof search for \IEL\, can be reduced to the case of so-called minimal derivations
(Section \ref{SEC7}). We implement it as a game of polynomial complexity and use the
characterization AP=PSPACE (see \cite{ChKoSt}) to prove the upper complexity bound for
\IEL. The matching lower bound follows from the same bound for intuitionistic
propositional logic \cite{Stat}.

\section{Sequent formulation of \IEL}
The definition of intuitionistic sequents is standard (see \cite{TrSh}). Formulas are
build from propositional variables and $\bot$ using $\wedge$, $\vee$, $\to$ and $K$;
$\neg F$ means $F\to\bot$. A sequent has the form $\Gamma\To F$ where $F$ is a formula
and $\Gamma$ is a multiset of formulas. $K(\Gamma)$ denotes $KF_1,\ldots,KF_n$ when
$\Gamma=F_1,\ldots,F_n$.

Let $\IELG$ be the extension of the intuitionistic propositional sequent calculus (e.g.
the propositional part of {\sf G2i} from \cite{TrSh} with the cut-rule) by the following
modal rules:

$$
\Gamma,K\bot\Rightarrow F\,, \qquad \ruleone{\Gamma\Rightarrow F}{K(\Gamma)\Rightarrow
KF\using{(KI_0)}}\,, \qquad \ruleone{\Gamma,F,KF\Rightarrow G}{\Gamma,F\Rightarrow
G\using{(KC)}}\,.
$$

\com $\IELG$ is a straightforwardly formulated sequent counterpart of \IEL. Instead of
the $K$-contraction rule $(KC)$ one can take the equivalent $K$-elimination rule:
$$
\ruleone{\Gamma,KF\Rightarrow G}{\Gamma,F\Rightarrow G\using{(KE)}}\,.
$$

\begin{theo}\label{IELG}
$\IELG\vdash \Gamma\Rightarrow F$ iff \ \ ${\sf IEL}\vdash \wedge\Gamma\rightarrow F$.
\end{theo}
\proof Straightforward induction on the derivations.

\fin

Our goal is to eliminate the cut-rule. But the cut-elimination result for $\IELG$ will
not have the desirable consequences, namely,  the subformula property and termination of
the proof search procedure. Below we give a different formulation without these
disadvantages.

\section{Cut-free variant $\IELGm$ with rules $(KI_1)$ and $(U)$}\label{SEC3}

Axioms:
$$
\Gamma,A\To A, \quad\mbox{$A$ is a variable or $\bot$.}
$$

\noindent Rules:
$$
\begin{array}{cc}
\ruleone{\Gamma,F,G\To H}{\Gamma,F\wedge G\To H\using{(\wedge \To )}} &
\ruletwo{\Gamma\To F}{\Gamma\To G}{\Gamma\To F\wedge G\using{(\To\wedge)}}\\[12pt]
\ruletwo{\Gamma,F\To H}{\Gamma,G\To H}{\Gamma,F\vee G\To H\using{(\vee\To)}}&
\ruleone{\Gamma\To F_i}{\Gamma\To F_1\vee F_2\using{(\To\vee )_i\quad (i=1,2)}}\\[12pt]
\ruletwo{\Gamma,F\rightarrow G\To F}{\Gamma,G\To H}{\Gamma,F\rightarrow G\To
H\using{(\rightarrow\To)}}
&\ruleone{\Gamma,F\To G}{\Gamma\To F\rightarrow G\using{(\To\rightarrow)}}\\[12pt]
\ruleone{\Gamma,K(\Delta),\Delta\To F} {\Gamma,K(\Delta)\To KF\using{(KI_1)}} &
\ruleone{\Gamma\To K\bot}{\Gamma\To F\using{(U)}}
%\ruleone{\Gamma\To K\bot}{\Gamma\To \bot\using{(\To\bot)}}
\end{array}
$$
In the rule $(KI_1)$ we additionally require that $\Gamma$ does not contain formulas of
the form $KG$. (This requirement is unessential, see Corollary \ref{ExtKI}.)

We define the main (occurrences of) formulas for axioms and for all inference rules
except $(KI_1)$ as usual --- they are the displayed formulas in the conclusions (not
members of $\Gamma,H$). For the rule $(KI_1)$ all members of $K(\Delta)$ and the formula
$KF$ are main.

\com In $\IELGm$ we do not add $(KE)$ or $(KC)$, but modify $(KI_0)$. In the presence of
weakening (it is admissible, see Lemma \ref{Weekening}) $(KI_0)$ is derivable:
$$
\ruleone {\ruleone{\Gamma\To F}{K(\Gamma),\Gamma\To F\using{(W)}}} {K(\Gamma)\To
KF\using{(KI_1)}}\,.
$$
So one can derive all sequents of the forms $F\To F$ for complex $F$ and $F\To KF$. The
latter also requires weakening in the case of $F=KG$:
$$
\ruleone{F\To F}{F\To KF\using{(KI_1),\;F\not=KG,}} \qquad\qquad \ruleone{ \ruleone
{KG\To KG} {KG,G\To KG\using{(W)}} }{KG\To KKG\using{(KI_1)}}\,.
$$

\com $(U)$ is necessary. There is no way to prove the sequent $K\bot\To \bot$  without
it.

\section{Structural rules are admissible}\label{SEC4}
We prove the admissibility of depth-preserving weakening and depth-preserving
contraction. Our proof follows \cite{TrSh} except the case of the rule $(KI_1)$. The
corresponding inductive step in the proof of Lemma \ref{Contraction} does not require
the inversion of the rule. Instead of it, some kind of contraction is build in the rule
itself.

We write $\vdash_n \Gamma\To F$ for ``$\Gamma\To F$ has a $\IELGm$-proof of depth at
most $n$''.

\begin{lem}[Depth-preserving weakening]\label{Weekening}
If \ $\vdash_n\Gamma\To F$ then $\vdash_n\Gamma, G\To F$.
\end{lem}
\proof Induction on $n$, see \cite{TrSh}. \fin

\begin{cor} \label{ExtKI} The extended $K$-introduction rule
$$
\ruleone{\Gamma_1,K(\Delta),\Delta,\Gamma_2\To F} {\Gamma_1,K(\Delta,\Gamma_2)\To
KF\using{(KI_{ext})}}
$$
is admissible in $\IELGm$ and \ $\vdash_n \Gamma_1,K(\Delta),\Delta,\Gamma_2\To F$
 implies \ $\vdash_{n+1}\Gamma_1,K(\Delta,\Gamma_2)\To KF$.
\end{cor}
\proof Suppose $\vdash_n\Gamma_1,K(\Delta),\Delta,\Gamma_2\To F$ and
$\Gamma_1=\Gamma_1',K(\Gamma_1'')$ where $\Gamma_1'$ does not contain formulas of the
form $KG$. By Lemma \ref{Weekening},
$$
\vdash_n\Gamma_1',K(\Gamma_1''),K(\Delta),K(\Gamma_2),\Gamma_1'',\Delta,\Gamma_2\To F\,.
$$
So,
$$
\vdash_{n+1}\Gamma_1',K(\Gamma_1'',\Delta,\Gamma_2)\To KF
$$
by $(KI_1)$. But $\Gamma_1',K(\Gamma_1'',\Delta,\Gamma_2)=\Gamma_1,K(\Delta,\Gamma_2)$,
so $\vdash_{n+1}\Gamma_1,K(\Delta,\Gamma_2)\To KF$.

\fin
\begin{cor}\label{AX}
All axioms of \ $\IELG$ are provable in $\IELGm$.
\end{cor}
\proof It is sufficient to prove sequents $\Gamma,\bot\To F$ and $\Gamma,K\bot\To F$:
$$
\ruleone{ \ruleone {\Gamma,\bot\To \bot}{\Gamma,\bot\To K\bot\using{(KI_{ext})}}
}{\Gamma,\bot\To F\using{(U)}}\,, \qquad\qquad \ruleone {\ruleone{\Gamma,\bot\To
\bot}{\Gamma,K\bot\To K\bot\using{(KI_{ext})}}} {\Gamma,K\bot\To F\using{(U)}}\,.
$$
\fin

\begin{lem}[Inversion lemma \cite{TrSh}]\label{INVERSION} Left rules are
invertible in the following sense:

If \ $\vdash_n \Gamma, A\wedge B\To C$ \ then \ $\vdash_n \Gamma, A, B\To C$.

If \ $\vdash_n \Gamma, A_1\vee A_2\To C$ \ then \ $\vdash_n \Gamma, A_i\To C$, $i=1,2$.

If \ $\vdash_n \Gamma, A\rightarrow B\To C$ \ then \ $\vdash_n \Gamma, B\To C$.
\end{lem}

\begin{lem}[Depth-preserving contraction]\label{Contraction}
If \ $\vdash_n\Gamma,F,F\To G$ \ then \ $\vdash_n \Gamma,F\To G$.
\end{lem}
\proof  Induction on $n$. Case $n=1$. When the first sequent is an axiom, the second one
is an axiom too.

Case $n+1$. When the displayed two occurrences of $F$ in $\Gamma,F,F\To G$ are not main
for the last rule of the derivation, apply the induction hypothesis to the premises of
the rule and contract $F$ there.

Suppose one of the occurrences is main. Only axioms may have atomic main formulas, so we
treat atomic $F$ as in case $n=1$.

When $F$ has one of the forms $A\wedge B$, $A\vee B$ or $A\to B$, we use the same proof
as in \cite{TrSh}. It is based on the items of Inversion lemma formulated in Lemma
\ref{INVERSION}.

Case $F=KA$ is new. The derivation of $\Gamma,F,F\To G$ of depth $n+1$ has the form
$$
\ruleone{\ruleone{{\cal D}}{\Gamma', K(\Delta),\Delta\To B}} {\Gamma', K(\Delta)\To
KB\using{(KI_1)}}
$$
where $\Gamma,F,F=\Gamma', K(\Delta)$ and $G=KB$; the multiset $\Delta$ contains two
copies of $A$. We have
\begin{equation}\label{EQ1}
\vdash_n \Gamma', K(\Delta),\Delta\To B.
\end{equation}

Let $(\,)^-$ means to remove one copy of $A$ from a multiset. We apply the induction
hypothesis to (\ref{EQ1}) and obtain $\vdash_n \Gamma, K(\Delta^-),\Delta^-\To B$. Then,
by $(KI_1)$,
$$
\vdash_{n+1}\Gamma, K(\Delta^-)\To KB\,.
$$
But \ $\Gamma,F=\Gamma', K(\Delta^-)$, so \ $\vdash_{n+1} \Gamma,F\To G$.

\fin

\section{Admissible modal rules}\label{SEC5}
We have already seen that $(KI_0)$ is admissible in $\IELGm$.

\begin{lem}[Depth-preserving $K$-elimination]\label{RuleKE}
If \ $\vdash_n\Gamma,KF\To G$ then \ $\vdash_n\Gamma,F\To G$.
\end{lem}
\proof Induction on $n$. Case $n=1$. When the first sequent is an axiom, the second one
is an axiom too.

Case $n+1$. Consider a proof of depth $n+1$ of a sequent $\Gamma,KF\To G$. Let $(R)$ be
its last rule. When the displayed occurrence of $KF$ is not main for $(R)$, apply the
induction hypothesis to its premises and then apply $(R)$ to reduced premises. It will
give $\vdash_n\Gamma,F\To G$.

Suppose the occurrence of $KF$ is main. The derivation has the form
$$
\ruleone{\vdash_n\Gamma',K(\Delta),KF,\Delta,F\To G'}
{\vdash_{n+1}\Gamma',K(\Delta,F)\To KG'\using{(KI_1)}} \,.
$$
Apply the induction hypothesis to the premise and remove one copy of $F$.  By Lemma
\ref{Contraction}\,, \ $\vdash_n\Gamma',K(\Delta),\Delta,F\To G'$. Then apply an
instance of $(KI_{ext})$ with $\Gamma_1=\Gamma',F$ and empty $\Gamma_2$. By Corollary
\ref{ExtKI}\,, \ $\vdash_{n+1}\Gamma',K(\Delta),F\To KG'$.

\fin

\begin{cor}[Depth-preserving $K$-contraction] \label{RuleKC}
If $\vdash_n\Gamma,KF,F\To G$ then $\vdash_n\Gamma,F\To G$.
\end{cor}
\proof Apply $(KE)$ and contraction. Both rules are admissible and preserve the depth
(Lemmas \ref{RuleKE}, \ref{Contraction}). \fin

\section{Cut is admissible}\label{SEC6}

Consider an $\IELGm$-derivation with additional cut-rule
\begin{equation}\label{CutRule}
\ruletwo{\Gamma_1\To F}{\Gamma_2,F\To G}{\Gamma_1,\Gamma_2\To G\using{(Cut)}}\,.
\end{equation}
and some instance of $(Cut)$ in it. The level of the cut is the sum of the depths of its
premises. The rank of the cut is the length of $F$.

\begin{lem}\label{CutELIM}
Suppose the premises of $(Cut)$ are provable in $\IELGm$ without $(Cut)$. Then the
conclusion is also provable in $\IELGm$ without $(Cut)$.
\end{lem}

\proof We define the following well-ordering on pairs of natural numbers:
$(k_1,l_1)>(k_2,l_2)$ iff $k_1>k_2$ or $k_1=k_2$ and $l_1>l_2$ simultaneously. By
induction on this  order we prove that a single cut of rank $k$ and level $l$ can be
eliminated.

As in \cite{TrSh}, we consider three possibilities:

I. One of the premises is an axiom. In this case the cut-rule can be eliminated. If the
left premise of (\ref{CutRule}) is an axiom,
$$
\ruletwo{\Gamma_1',A\To A}{\Gamma_2,A\To G}{\Gamma_1',A,\Gamma_2\To G\using{(Cut)}}\,,
$$
then $(Cut)$ is unnecessary. The conclusion can be derived from the right premise by
weakening (Lemma \ref{Weekening}).

Now suppose that the right premise is an axiom. If the cutformula $F$ is not main for
the axiom $\Gamma_2,F\To G$ then the conclusion $\Gamma_1,\Gamma_2\To G$ is also an
axiom, so $(Cut)$ can be eliminated. If $F$ is main for the right premise then $F=G=A$
where $A$ is atomic, so (\ref{CutRule}) has the form
$$
\ruletwo {\Gamma_1\To A} {\Gamma_2,A\To A}{\Gamma_1,\Gamma_2\To A\using{(Cut)}}\,.
$$
The conclusion can be derived without $(Cut)$ from the left premise by weakening (Lemma
\ref{Weekening}).

II. Both premises are not axioms and the cutformula is not main for the last rule in the
derivation of at least one of the premises. In this case one can permute the cut upward
and reduce the level of the cut. The cutformula remains the same, so the cut rule can be
eliminated by induction hypothesis (see \cite{TrSh}).

III. The cutformula $F$ is main for the last rules in the derivations of both premises.
In this case we reduce the rank of cut and apply the induction hypothesis.

Note that $F$ is not atomic. (The atomic case is considered in I.) If the last rule in
the derivation of the left premise is $(U)$ then $(Cut)$ can be eliminated:
$$
\ruletwo {\ruleone{\Gamma_1\To K\bot}{\Gamma_1\To F\using{(U)}}} {\Gamma_2,F\To
G}{\Gamma_1,\Gamma_2\To G\using{(Cut)}} \quad \rightsquigarrow\quad
\ruleone{\Gamma_1,\Gamma_2\To K\bot}{\Gamma_1,\Gamma_2\To G\using{(U)}}\,.
$$

Case $F=KA$, the last rule in the derivation of the left premise is $(KI_1)$:
$$
\ruletwo { \ruleone{\Gamma,K(\Delta),\Delta\To A} {\Gamma,K(\Delta)\To KA
\using{(KI_1)}} } { \ruleone{\ruleone{{\cal D}'}{\Gamma',K(\Delta',A),\Delta',A\To B}}
{\Gamma',K(\Delta'),KA\To KB\using{(KI_1)}} } {\Gamma,K(\Delta),\Gamma',K(\Delta') \To
KB\using{(Cut)}} \quad\rightsquigarrow
$$
From $\Gamma',K(\Delta',A),\Delta',A\To B$ by $K$-contraction (Corollary \ref{RuleKC})
we obtain $\Gamma',K(\Delta'),\Delta',A\To B$ and then reduce the rank:

$$
\rightsquigarrow\qquad \ruleone {\ruletwo { \Gamma,K(\Delta),\Delta\To A } {
\ruleone{{\cal D}''} {\Gamma',K(\Delta'),\Delta',A\To B} }
{\Gamma,K(\Delta),\Delta,\Gamma',K(\Delta'),\Delta' \To B \using{(Cut)}}}
{\Gamma,K(\Delta),\Gamma_1',K(\Delta') \To KB \using{(KI_1)}} \,.
$$

In remaining cases (when $F$ has one of the forms $A\wedge B$, $A\vee B$ or $A\to B$) we
follow \cite{TrSh}. \fin

\begin{theo}\label{CutTheorem}  $(Cut)$ is admissible in $\IELGm$.
\end{theo}
\proof It is a consequence of Lemma \ref{CutELIM}.

\fin

\com Our formulation of the rule $(KI_1)$ combines K-introduction with contraction. It
is done in order to eliminate the contraction rule and to avoid the case of contraction
in the proof of Lemma \ref{CutELIM}. But the contraction rule remains admissible and can
be added as a ground rule too, so  we can simplify the formulation of the
$K$-introduction rule. It results in a ``light'' cut-free version $\IELGp$:

\medskip
Axioms:  \quad $\Gamma,A\To A$,\qquad  $A$ is a variable or $\bot$.

Rules:
$$
\ruleone{\Gamma,\Delta,\Delta\To G}{\Gamma,\Delta\To G\using{(C)}}
$$
$$
\begin{array}{cc}
\ruleone{\Gamma,F,G\To H}{\Gamma,F\wedge G\To H\using{(\wedge \To )}} &
\ruletwo{\Gamma\To F}{\Gamma\To G}{\Gamma\To F\wedge G\using{(\To\wedge)}}\\[12pt]
\ruletwo{\Gamma,F\To H}{\Gamma,G\To H}{\Gamma,F\vee G\To H\using{(\vee\To)}}&
\ruleone{\Gamma\To F_i}{\Gamma\To F_1\vee F_2\using{(\To\vee )_i\quad (i=1,2)}}\\[12pt]
\ruletwo{\Gamma\To F}{\Gamma,G\To H}{\Gamma,F\rightarrow G\To H\using{(\rightarrow\To)}}
&\ruleone{\Gamma,F\To G}{\Gamma\To F\rightarrow G\using{(\To\rightarrow)}}\\[12pt]
\ruleone{\Gamma_1,\Gamma_2\To F} {\Gamma_1,K(\Gamma_2)\To KF\using{(KI)}} &
\ruleone{\Gamma\To K\bot}{\Gamma\To F\using{(U)}}
\end{array}
$$

\begin{lem}\label{EquivIELG}
$\IELGp\vdash \Gamma\To F$ \ iff \ $\IELGm\vdash \Gamma\To F$.
\end{lem}
\proof Part ``only if''. The rule $(KI)$ is a particular case of $(KI_{ext})$, so all
rules of $\IELGp$ are admissible in $\IELGm$ (Lemmas \ref{Contraction}, \ref{Weekening}
and Corollary \ref{ExtKI}).

Part ``if''. All missing rules are derivable in $\IELGp$:
$$
\ruleone {\ruletwo{\Gamma,F\rightarrow G\To F}{\Gamma,G\To H} {\Gamma,F\rightarrow
G,F\rightarrow G\To H\using{(\rightarrow\To)}}} {\Gamma,F\rightarrow G\To H\using{(C)}}
\,,\qquad \ruleone {\ruleone{\Gamma,K(\Delta),\Delta\To F}
{\Gamma,K(\Delta),K(\Delta)\To KF\using{(KI)}}} {\Gamma_1,K(\Delta)\To KF\using{(C)}}\,.
$$
\fin

\begin{theo}
$(Cut)$ is admissible in $\IELGp$.
\end{theo}
\proof Lemma \ref{CutELIM} implies the similar statement for the calculus $\IELGp$.
Indeed, one can convert $\IELGp$-derivations into $\IELGm$-derivations, eliminate a
single cut in $\IELGm$, and then convert the cut-free $\IELGm$-derivation backward
(Lemma \ref{EquivIELG}). The statement implies the theorem. \fin

\begin{theo}\label{EQUIVCALC} The following are equivalent:\\[3pt]
$
\begin{array}{ll}
1. & \IELG\vdash\Gamma\To F .\\
2. & \IELGm\vdash\Gamma\To F. \\
3. & \IELGp\vdash\Gamma\To F .\\
4. & {\sf IEL}\vdash\wedge\Gamma\rightarrow F.
\end{array}
$
\end{theo}
\proof $1.\Leftrightarrow 2.$ All rules of $\IELG$ are admissible in $\IELGm$ (Lemmas
\ref{Weekening}, \ref{Contraction}, \ref{RuleKE}, Theorem \ref{CutTheorem}) and vice
versa.

The equivalence of 2. and 3. is proved in Lemma \ref{EquivIELG}, the equivalence of 1.
and 4. -- see Theorem \ref{IELG}. \fin

\section{Complexity of  \IEL}\label{SEC7}

We prove that \IEL \ is PSPACE-complete. The lower bound follows from the same lower
bound for the intuitionistic propositional logic. To prove the upper bound we show that
polynomial space is sufficient for the proof search. Our proof search technique  is
based on monotone derivations and is similar to one used in \cite{NVK}.\footnote{In
\cite{NVK} the definition of a monotone derivation contains a missprint, but actually
the correct Definition \ref{DEFMON} is used. }

\begin{defin}\label{DEFMON}\rm
For a multiset $\Gamma$ let  $\mbox{\it set}(\Gamma)$ be the set of all its members. An
instance of a rule
$$
\ruleone{\Gamma_1\Rightarrow F_1\;\ldots\;\Gamma_n\Rightarrow F_n } {\Gamma\Rightarrow
F}
$$
is {\em monotone} if ${\displaystyle \mbox{\it set}(\Gamma)\subseteq \bigcap_i \mbox{\it
set}(\Gamma_i)}$. A derivation is called monotone if it uses monotone instances of
inference rules only.
\end{defin}

Consider the extension $\IELGpp$ of the calculus $\IELGm$ by the following rules: the
contraction rule  $(C)$ and
$$
\begin{array}{cc}
\ruleone{\Gamma,F\wedge G,F,G\To H} {\Gamma,F\wedge G,F\To H\using{(\wedge_1^C\To )}}\,,
& \ruleone{\Gamma,F\wedge G,F,G\To H}
{\Gamma,F\wedge G,G\To H\using{(\wedge_2^C\To )}}\,, \\[12pt]
\ruleone{\Gamma,F\wedge G,F,G\To H}{\Gamma,F\wedge G\To H\using{(\wedge^C \To )}}\,, &
\ruletwo{\Gamma,F\vee G,F\To H}{\Gamma,F\vee G,G\To H}{\Gamma,F\vee G\To H\using{(\vee^C\To)}}\,, \\[12pt]
\ruleone{\Gamma,F\To G}{\Gamma,F\To F\rightarrow G\using{(\To\rightarrow^W)}}\,, &
\ruletwo{\Gamma,F\rightarrow G\To F}{\Gamma,F\rightarrow G,G\To H}{\Gamma,F\rightarrow
G\To H\using{(\rightarrow^C\To)}}\,,
\end{array}
$$
$$
\ruleone{\Gamma,K(\Delta_1,\Delta_2),\Delta_1,\Delta_2\To F}
{\Gamma,\Delta_1,K(\Delta_1,\Delta_2)\To KF\using{(KI_1^W)}}\,.
$$
In $(KI_1^W)$ we require that the multiset $\Gamma,\Delta_1$ does not contain formulas
of the form  $KG$.

\begin{lem}\label{EQUIVPP}
$\IELGpp\vdash \Gamma\To F$ \ iff \ $\IELGm\vdash \Gamma\To F$.
\end{lem}
\proof All new rules are  some combinations of corresponding ground rules with
structural rules. The latter are admissible in $\IELGm$ (Lemmas \ref{Contraction},
\ref{Weekening}). \fin

\begin{lem}\label{MON}
Any derivation in $\IELGpp$ can be converted into a monotone derivation of the same
sequent.
\end{lem}

\proof Consider a derivation which is not monotone. Chose the first non-monotone
instance $(R)$ of a rule in it. $(R)$ introduces a new formula $A$ in the antecedent of
its conclusion which is not present in antecedents of some of its premises. Add a copy
of $A$ to the antecedent of the conclusion and to antecedents of all sequents above it.
When $A$ has the form $KB$ and is added to the antecedent of the conclusion of some
instance of rules $(KI_1)$ or $(KI_1^W)$ above $(R)$, add a copy of $B$ to the
antecedent of the premise of this rule and to antecedents of all sequents above it. When
$B$ has the form $KC$, do the same with $C$, etc. Finally, insert the contraction rule
after $(R)$:
$$
\ruleone{{\cal D}}{A,\,\Gamma\Rightarrow F\using{(R)}} \qquad\rightsquigarrow\qquad
\ruleone{\ruleone{{\cal D'}}{A,\,A,\,\Gamma\Rightarrow F\using{(R)}}}
{A,\,\Gamma\Rightarrow F\using(C)} \,.
$$

The result is also a correct derivation with one non-monotone instance eliminated.
Repeat the transformation until the derivation becomes monotone.

\fin

\begin{lem}\label{MIN}
A monotone derivation of a sequent $\Gamma\To F$ in $\IELGpp$  can be converted into a
monotone derivation of the sequent \ ${\it set}(\Gamma)\To F$ that contains only
sequents of the form  \ ${\it set}(\Gamma')\To F'$. The transformation does not increase
the depth of the proof.
\end{lem}

\proof Given a monotone derivation replace all sequents $\Gamma'\To F'$ in it with ${\it
set}(\Gamma')\To F'$. This transformation converts axioms into axioms. We claim that an
instance of an inference rule will be converted either into some other instance of a
rule of $\IELGpp$  or some premise of the converted instance will coincide with its
conclusion, so the rule can be removed from the resulting proof. The depth of the proof
does not increase.

Indeed, instances of $(\To\wedge)$, $(\To\vee)$ and $(U)$ will be converted into some
other instances of the same rule. An instance of $(C)$ will be converted into the
trivial rule that can be removed:
$$
\ruleone{\Gamma,\Delta,\Delta\To G}{\Gamma,\Delta\To G\using{(C)}}
\qquad\rightsquigarrow\qquad \ruleone{set(\Gamma,\Delta)\To G}{set(\Gamma,\Delta)\To G}
\qquad\rightsquigarrow\qquad \mbox{remove.}
$$

The remaining cases. Let $k,l,m,n,k',l',m',n'\geq 0$ and
$F^k=\underbrace{F,\ldots,F}_{k\mbox{ à §}}$.

\noindent All monotone instances of $(\wedge\To)$, $(\wedge_1^C\To)$, $(\wedge_2^C\To)$,
$(\wedge^C\To)$ have the form
$$
\ruleone{\Gamma,(F\wedge G)^{k+1},F^{l+1},G^{m+1}\To H} {\Gamma,(F\wedge
G)^{k'+1},F^{l'},G^{m'}\To H}\,.
$$
Contractions in antecedents will give
$$
\begin{array}{ll}
\ruleone{\Gamma',F\wedge G,F,G\To H}
{\Gamma',F\wedge G\To H\using{(\wedge^C\To )}}, &  l'=m'=0,\\[12pt]
\ruleone{\Gamma',F\wedge G,F,G\To H} {\Gamma',F\wedge G,F\To H\using{(\wedge_1^C\To )}},
&  l'>0,m'=0, \\ [12pt] \ruleone{\Gamma',F\wedge G,F,G\To H} {\Gamma',F\wedge G,G\To
H\using{(\wedge_2^C\To )}}, &  l'=0,m'>0,\\ [12pt] \mbox{trivial rule (remuved)}, &
l',m'>0.
\end{array}
$$
All monotone instances of  $(\vee\To)$, $(\vee^C\To)$ have the form
$$
\ruletwo{\Gamma,(F\vee G)^{k+1},F^{l+1},G^{m}\To H} {\Gamma,(F\vee
G)^{k+1},F^{l},G^{m+1}\To H} {\Gamma,(F\vee G)^{k'+1},F^{l},G^{m}\To H}\,.
$$
Contractions in antecedents will give
$$
\begin{array}{ll}
\ruletwo{\Gamma',F\vee G,F\To H}{\Gamma,F\vee G,G\To H}
{\Gamma',F\vee G\To H\using{(\vee^C\To)}}, &  l=m=0,\\[12pt]
\mbox{trivial rule (remuved)}, &   l>0\mbox{ or } m>0.
\end{array}
$$
All monotone instances of $(\To\rightarrow)$, $(\To\rightarrow^W)$ have the form
$$
\ruleone{\Gamma,F^{k+1}\To G} {\Gamma,F^{k'}\To F\rightarrow G}\,.
$$
Contractions in antecedents will give
$$
\begin{array}{ll}
\ruleone{\Gamma',F\To G}
{\Gamma',F\To F\rightarrow G\using{(\To\rightarrow^W)}}, &  k'>0, \\[12pt]
\ruleone{\Gamma',F\To G} {\Gamma'\To F\rightarrow G\using{(\To\rightarrow)}}, &  k'=0.
\end{array}
$$
All monotone instances of $(\rightarrow\To)$, $(\rightarrow^C\To)$ have the form
$$
\ruletwo{\Gamma,(F\rightarrow G)^{k+1},G^{l}\To F} {\Gamma,(F\rightarrow
G)^{k'+1},G^{l+1}\To H} {\Gamma,(F\rightarrow G)^{k+1},G^{l}\To H}.
$$
Contractions in antecedents will give
$$
\begin{array}{ll}
\ruletwo{\Gamma',(F\rightarrow G)\To F} {\Gamma',(F\rightarrow G),G\To H}
{\Gamma',(F\rightarrow G)\To H\using{(\rightarrow^C\To)}},
& l=0, \\[12pt]
\mbox{trivial rule (remuved)}, &   l>0.
\end{array}
$$
All monotone instances of $(KI_1)$, $(KI_1^W)$ have the form
$$
\ruleone{\Gamma,G^{k+1},(KG)^{l+1},\ldots,H^{m+1},(KH)^{n+1}\To F}
{\Gamma,G^{k'},(KG)^{l'+1},\ldots,H^{m'},(KH)^{n'+1}\To KF}
$$
Contractions in antecedents will give
$$
\begin{array}{ll}
\ruleone{\Gamma',G,KG,\ldots,H,KH\To F} {\Gamma',KG,\ldots,KH\To KF\using{(KI_1)}},
&  k'=0,\ldots,m'=0, \\[12pt]
\ruleone{\Gamma',G,KG,\ldots,H,KH\To F} {\Gamma',KG,\ldots,H,KH\To KF\using{(KI_1^W)}},
&  k'=0,\ldots,m'>0, \\[12pt]
\mbox{trivial rule (remuved)}, &   k',\ldots,m'>0.\end{array}
$$
\fin

\begin{lem}[Subformula property]\label{SUBF}
Consider a derivation of a sequent $\Gamma\To F$ in $\IELGm$, $\IELGp$ or $\IELGpp$. Any
sequent in it is composed of subformulas of some formulas from the multiset \ $\Gamma,
F, K\bot$.
\end{lem}
\proof For any rule of these calculi, its premises are composed of subformulas of
formulas occuring in its conclusion and, possibly, of $K\bot$.
 \fin

\begin{defin}\rm A monotone $\IELGpp$-derivation of a sequent \ ${\it set}(\Gamma)\To F$
is called minimal if it contains only sequents of the form  \ ${\it set}(\Gamma')\To F'$
and has the minimal depth.

The size of a sequent $F_1,\ldots ,F_k\To F$ is the sum of the lengths of all formulas
$F_i$ and $F$.
\end{defin}

\begin{lem} \label{QGAME}
Let ${\cal M}_n$ be the set of all minimal derivations of sequents of size $n$. There
exist polynomials $p$ and $q$ such that for any derivation ${\cal D}\in {\cal M}_n$,
its depth is bounded by $p(n)$ and the sizes of all sequents in ${\cal D}$  do not
exceed $q(n)$.
\end{lem}

\proof Consider a proof tree for some ${\cal D}\in {\cal M}_n$ and a path from the root
to some leaf in it:

$$
\Gamma_0\Rightarrow F_0, \ldots , \Gamma_N\Rightarrow F_N.
$$
All sequents in it are distinct from each other, all of them composed of subformulas of
the first sequent, $\bot$ and $K\bot$ (Lemma \ref{SUBF}), and
$\Gamma_i\subseteq\Gamma_{i+1}$ holds for $i<N$.

Divide the path into maximal intervals with the same $\Gamma_i$ inside. The length of
such interval is bounded by the number of possible formulas $F_i$, which is $O(n)$. The
number of intervals is $O(n)$ too, because it does not exceed the maximal length of a
strictly monotone sequence $\Delta_0\subset\Delta_1\subset\ldots\subset\Delta_k$ of
subsets of $S$ where $S$ is the set of all subformulas of the first sequent extended by
$\bot$ and $K\bot$. So, $|S|=O(n)$ and $N=O(n^2)$.

Any sequent $\Gamma_i\To F_i$ consists of at most $|S|+1$ formulas of length $O(n)$, so
its size is $O(n^2)$. \fin

\begin{cor} \label{PSPACE}
The set of all \ $\IELGpp$-derivable sequents belongs to ${\sf PSPACE}$.
\end{cor}

\proof The result follows from the known game characterization ${\sf AP=PSPACE}$
(\cite{ChKoSt}, see also \cite{VKru} or \cite{KiShV}).

 Let $p$, $q$ be polynomials from Lemma \ref{QGAME}.
Consider the following two-person game with players $(P)$ and $(V)$. The initial
configuration $b_0$ is a sequent of the form ${\it set}(\Gamma)\Rightarrow F$ of size
$n$. Player $(P)$ moves the first. He writes down one or two sequents of sizes less than
$q(n)$ and his opponent $(V)$ chooses one of them, and so on.  The game is over after
$p(n)$ moves of $(V)$ or when $(V)$ chooses a sequent that is an axiom of $\IELGpp$.

Let $w_i$ and $b_i$ denote the moves of players $(P)$ and $(V)$ respectively, so $b_0,
w_1, b_1, b_2, w_2,...$ \ is a run of the game. Player $(P)$ wins if the following
conditions are satisfied:
\begin{enumerate}
\item For every move of $(P)$  the figure \ $\ruleone{w_{i}}{b_{i-1}}$ \ is a monotone
instance of some inference rule of $\IELGpp$.

\item All sequents written by $(P)$ have the form ${\it set}(\Delta)\Rightarrow G$.

\item At his last move $(V)$ is forced to choose an axiom of $\IELGpp$.
\end{enumerate}

The number and the sizes of moves are bounded by polynomials and the winning condition
is polynomial-time decidable, so the set  $M$ of initial configurations that admit a
winning strategy for $(P)$ belongs to ${\sf PSPACE}$ (see \cite{ChKoSt}).

By Lemma \ref{QGAME}, a sequent belongs to $M$ iff it has a minimal derivation. But it
follows from Lemmas \ref{MON}, \ref{MIN}, \ref{Weekening}, that a sequent $\Gamma\To F$
is $\IELGpp$-derivable iff ${\it set}(\Gamma)\Rightarrow F$ has a minimal derivation.
Thus, the general derivability problem for $\IELGpp$ belongs to ${\sf PSPACE}$ too.
 \fin

\begin{theo}\label{MAINPSPACE}
The derivability problems for $\IELG$, $\IELGm$, $\IELGp$, $\IELGpp$ and {\sf IEL} are
$\sf PSPACE$-complete.
\end{theo}
\proof The lower bound $\sf PSPACE$ follows from the same lower bound for intuitionistic
propositional logic \cite{Stat}. The upper bound $\sf PSPACE$ for $\IELGpp$ is
established in Corollary \ref{PSPACE}. It can be extended to other calculi by Theorem
\ref{EQUIVCALC} and Lemma \ref{EQUIVPP}. \fin

\section*{Acknowledgements}

We are grateful to Sergey Artemov for inspiring discussions of intuitionistic epistemic
logic.

The research described in this paper was partially supported by Russian Foundation for
Basic Research (grant 14-01-00127).

\end{document}